\documentclass[floats,aps,epsfig,amsmath,twocolumn]{revtex4}
\usepackage{epsfig}
\usepackage{amsmath}

\begin{document}


\title{Properties of dense partially random graphs}

\author{Sebasti\'{a}n Risau-Gusman}
\email{srisau@if.ufrgs.br} \affiliation{Instituto de F\'{\i}sica,
UFRGS, Caixa Postal 15051, 91501-970 Porto Alegre, RS, Brazil}

\date{\today}

\begin{abstract}
\begin{center}
\parbox{14cm}{We study the properties of random graphs where for each vertex a
{\it neighbourhood} has been previously defined. The probability
of an edge joining two vertices depends on whether the vertices are
neighbours or not, as happens in Small World Graphs (SWGs). But
we consider the case where the average degree of each node is of
order of the size of the graph (unlike SWGs, which are sparse).
This allows us to calculate the mean distance and clustering,
that are qualitatively similar (although not in such a dramatic
scale range) to the case of SWGs. We also obtain analytically the
distribution of eigenvalues of the corresponding adjacency
matrices. This distribution is discrete for large eigenvalues and
continuous for small eigenvalues. The continuous part of the
distribution follows a semicircle law, whose width is
proportional to the "disorder" of the graph, whereas the discrete
part is simply a rescaling of the spectrum of the substrate. We
apply our results to the calculation of the mixing rate and the
synchronizability threshold.}
\end{center}
\end{abstract}

\maketitle

\section{Introduction} \label{sec:intro}
Many natural and artificial systems are composed of a large number
of identical agents that interact. Examples of this kind of
systems are numerous and widespread: collaborators in physics,
neural networks, computer programs (considered as a system of
interacting subroutines), the World Wide Web, movie actors, etc.

The interactions however need not be identical and can vary in
strength and range. A simplified analysis of such a system can
take into account only the pattern of interactions, abstracting
everything else. What is left is usually called a {\it network}
and is mathematically represented by a {\it graph} (along this
article both terms will be used interchangeably.)

A graph is composed of {\it vertices} (associated to the agents)
connected by {\it edges}. Agents only interact with other agents
if there is an edge joining the corresponding vertices, and the
strength of the interaction is given by the {\it weight} of the
edge. The interactions between graph theory and physical science,
particularly physics, has been very fruitful~\cite{wu}.

In 1959 Erd\"{o}s and Renyi~\cite{ER} started a whole new branch
of Graph Theory by creating (and extensively studying) the concept
of {\it random graphs}. These are graphs where each edge has a
defined probability of being present, and this probability is
independent of all other edges. They seem particularly well
suited for the study of systems for which there is little
information about the range of the interactions. In these cases it
seems natural to assign independent and equal probabilities to
the different connections.

The average distance of a graph is defined as the average of the
length (i. e. the number of vertices) of the smallest path joining
two edges. The clustering coefficient gives the average number of
connections present between neighbors of a vertex, divided by the
number of possible connections within the neighborhood of the
vertex. In their groundbreaking article of 1998, Watts and
Strogatz~\cite{WS} (hereafter WS) showed that most real networks
display a very short average distance combined with large values
of the clustering coefficient. But random graphs do not fulfill
these requirements because, even though the average distance is
short, the clustering coefficient turns out to be rather small.
Thus they proposed a new graph model, the small world graph
(hereafter SWG). To build a SWG one starts with a regular and
{\it sparse} substrate graph (with large clustering but also with
large average distance), and then rewires some edges with a
probability $p$, thus creating shortcuts. It was shown that a
small number of shortcuts is enough to significantly lower the
average distance while leaving the clustering coefficient almost
unchanged.

Since their proposal, the properties of SWGs have been intensively
studied, with numerical as well as analytical methods. Even
though some numerical analysis have been performed~\cite{farkas},
one of the properties that still resists analytical treatment is
the spectrum of the adjacency matrix of SWGs.

The constraint of sparsity is justified by the fact that many
networks in nature display this characteristic. But one can also
find systems where the pattern of connections of every node spans
a significant portion of the whole graph (i.e. the degree of the
nodes is of the same order as the size of the network). Some
examples of this are the network of train routes in
India~\cite{sen}, the full reaction graphs of the metabolic
network of {\it E. Coli}~\cite{WF,LK} and the network of the
interacting units of the computer program Mozilla~\cite{moura}.

In this article we study the properties of graphs for which the
sparsity constraint has been dropped. This allows us to calculate
analytically the average distance and the clustering coefficient,
as well as the whole eigenvalue distribution of the corresponding
adjacency matrices. The spectra of these matrices can be
considered as limiting cases of the ones of sparse SWNs.

The values of the different properties calculated in this article
are only {\it averages} over a certain family of graphs (defined
in the next section). Nevertheless, simulations support the idea
that the properties of almost every graph of this family should
tend, in probability, to the average values found.

In Section~\ref{sec:model} we define the model and relate it to
the Watts-Strogatz prescription. In section~\ref{sec:metric} we
calculate the average distance and the clustering coefficient. In
section~\ref{sec:spectrum} we obtain the spectrum of the
corresponding adjacency matrices and compare it to the spectrum
of sparse graphs. A couple of applications of the results
obtained are presented in Section~\ref{sec:appli}. In
section~\ref{sec:conclu} some conclusions are drawn.

\section{The model}
\label{sec:model} A graph is a pair of sets $(V,E)$, where $V$
has $N$ elements (or vertices) and $E \subseteq V^2$ has $M$
elements (or edges). Two vertices $v_1$ and $v_2$ are connected
if $(v_1,v_2) \in E$.

WS~\cite{WS} proposed a graph model capable of interpolating
between order and randomness. The graphs in this model are built
by taking a {\it substrate}, which is a graph displaying some
regularity, and randomly {\it rewiring} some of its edges, by
keeping fixed one end of some edges and redirecting the other end
to a different vertex at random, but following some rule. Let us
consider, for example, the unidimensional case, where each vertex
in a ring is connected to $k/2$ vertices to the left, and $k/2$ vertices to
the right. The process begins by traveling clockwise on the ring,
and for each vertex rewiring, with probability $p$, the
connection that joins it with its first neighbor to the right.
Once the circle is completed, a new round is made where now the
connection rewired is the one to the second neighbor to the right.
The process ends after $k/2$ rounds (because $k/2$ is the number
of neighbors to the right). A similar process can be implemented
for a higher dimensional substrate. The graph obtained is called a
SWG.

In our model of dense partially random graphs (hereafter DPRGs),
we only consider hypercubic lattices as substrates, with the hope
that for other substrates things will not be very different, as
is the case for sparse SWG~\cite{watts}. In these lattices $V
\subset R^d$ and each dimension has a different connectivity
parameter $k_i$, which means that each node is connected to a
hypercube of $k=\prod_{i=1}^d k^i -1$ nodes. This defines a
$k$-neighborhood for each vertex. As opposed to the usual
constraint of considering sparse networks, here we are concerned
only with {\it dense} graphs, i. e. graphs where each vertex is
connected to $k=O(N)$ other vertices. Notice that, to have the
same number of neighbors for all nodes, we are considering
periodical boundary conditions for the lattices. Thus, the one
dimensional lattice is formed by points on a ring, the two
dimensional lattice by points on a torus, etc.

We randomize the graph in the following way: each edge in every
$k$-neighborhood is deleted with a probability $1-p_1$ and
vertices that do not belong to the same $k$-neighborhood are
joined with probability $p_2$. This is equivalent to saying that,
on an empty graph, each vertex is joined by a {\it short} link
with probability $p_1$ to every vertex in its $k$-neighborhood and
with probability $p_2$ to those outside it, by  a {\it long} link
or {\it shortcut}. A graph generated with this prescription is
called $G_{p_1 p_2}(\gamma)$ , with $\gamma=k/N$. The family of
all such graphs, for $p_1$, $p_2$ and $\gamma$ fixed is called
${\mathcal G}_{p_1 p_2}(\gamma)$. With $p_1=1$ and $p_2=0$ one
obtains the ordered substrate, whereas for $p_1=p_2=k/N$ one
obtains a random graph~\cite{bollobas}.

Notice that each vertex has a different number of connections,
whose average number is $k \, p_1+(N-1-k)p_2 \sim N \,(\gamma
p_1+(1-\gamma)p_2)$. But in the limit treated in this article, of
large values of $N$, the deviations from this average value are
exponentially small. To interpolate with only one parameter
between a fully ordered and a fully random graph, we fix this
average value by requiring that $\gamma p_1+(1-\gamma)
p_2=\gamma$ (along the paper, though, results will be displayed
for general $p_1$ and $p_2$, unless otherwise stated). Notice
that in this case the average number of connections of each
vertex is $k$. Choosing $p_1$ as the preferred parameter, a graph
satisfying the relationship shown above is called
$G_{p_1}(\gamma)$ and the family of graphs with $p_1$ and
$\gamma$ fixed is called ${\mathcal G}_{p_1}(\gamma)$. As
mentioned, for $p_1=\gamma$ the family ${\mathcal G}_{p_1}$
(\cite{bollobas}) of completely random graphs is obtained.

In our model the number of shortcuts is $p_2 N$, for large $N$.
To be able to compare the graphs in ${\mathcal G}_{p_1}(\gamma)$
with SWGs, it is necessary to know the number of shortcuts of the
graphs generated with the WS prescription. In the sparse case, it
is known that this number is $\sim pkN/2$. This comes from the
fact that the sparsity of the network ensures that when a link is
selected, the probability that it will be rewired to a vertex
inside the same $k$-neighbourhood is $ \sim 1/N$. But if the
network is dense, this probability becomes nonvanishing. The
following procedure provides a good approximation to the number
of shortcuts.

Instead of disconnect and rewire the links sequentially in each
round, let us assume that in each round $pN$ short edges are deleted {\it at
once}, and then $pN$ {\it random} edges are added, which can be
short or long. The process begins with only the $S=kN/2$ short
edges present in the graph. We call $S_t$ the number of short
edges after round $t$ and $L_t$ the number of long edges. Notice
that after deleting $pN$ short edges, the probability that one of
the random edges added is short is the quotient between the
number of available short edges and the number of total available
edges: $(pN+S-S_t)/(L+S-S_t-L_t+pN)=(pN+S-S_t)/(L+pN)$, where
$L=N(N-1)/2-S$. We are using the fact that the number of edges is
conserved ($S_t+L_t=S$). Using this, the evolution equations for
large enough $N$ can be obtained:

\begin{eqnarray}
\label{eq.A}
S_{t+1} &=& S_t -pN+ pN \frac{pN+S-S_t}{L+pN} \\
L_{t+1} &=& S-S_{t+1} ,
\end{eqnarray}

\noindent The second summand in Eq.~(\ref{eq.A}) corresponds to
the edges deleted and the third to the fraction of edges added
that are short. This formula can be iterated to obtain:

\begin{equation}
S_{t+1}=S-L+L\left(\frac{L}{L+pN} \right)^{t+1} .
\end{equation}

After $t=k/2$ rounds, and in the limit of large $k$ and $N$ such
that $\gamma=k/N$, we obtain:

\begin{eqnarray}
\label{eq.p1p2} p_1&=&S_k/S \sim  1-\frac{1-\gamma}{\gamma}
\left(1-\exp \left(-\frac{\gamma p}{1-\gamma} \right) \right)
\\
\label{eq.p2} p_2&=&L_k/L \sim 1-\exp \left(-\frac{\gamma
p}{1-\gamma} \right)
\end{eqnarray}

\begin{figure}
  \centerline{\epsfig{file=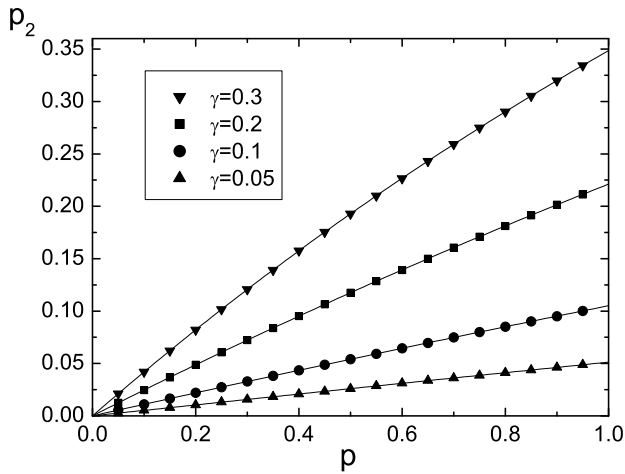,height=7cm}}
  \caption{Fraction of shortcuts as a function of the rewiring probability $p$. The symbols are averages
  taken on $30$ matrices with $N=6001$. The lines are the theoretical predictions. Error bars are smaller than the symbols.}
\label{fig.1}
\end{figure}

Figure~\ref{fig.1} shows that the agreement with the real number
of shortcuts for dense SWGs is very good. Thus, in what follows,
to translate the results of our model to dense small world graphs,
it suffices to take $p_2$ and invert Eq.~(\ref{eq.p2}) to obtain
the corresponding $p$. Notice that for values of $p$ close to 1
there is some "overshooting", as the resulting SWGs have more
shortcuts than the corresponding random graphs (i. e. $p_2>
\gamma$).

\section{Topological Properties} \label{sec:metric}
\subsection{Average Distance} \label{sec:avdist}

For random graphs with a number $O(N^2)$ of edges, it is known
that the diameter (i. e. the largest distance between any two
nodes) is equal to $2$, for large values of $N$. This means that
from any vertex only $1$ or $2$ steps are needed to reach any
other vertex. This in turn implies that the average distance in
such a graph tends to $2-p$. For a general graph it is
known~\cite{lovejoy} that $2-p$, with $p=2M/N(N+1)$ is a lower
bound to the average distance. It is interesting to notice that
this bound is achieved by some graphs (stars, for example).
Random graphs, on the other hand, only achieve it in the limit of
infinite $N$. In our model this implies that the average distance
satisfies ${\overline d} \geq 2-p_1 \gamma-p_2 (1-\gamma)$.

The graphs in ${\mathcal G}_{p_1 p_2}(\gamma)$ can be generated as
the union of two random graphs. One of them is simply a
completely random graph with edge probability $p_2$, called
$G_{p_2}$. For the other, the probability that an edge is present
is $0$ if it is a long edge and $p=\frac{p_1-p_2}{1-p_2}$ if it
is a short edge. This union results in a graph equivalent to the
one obtained by adding a number $N_a$ of edges to $G_{p_2}$, such
that ${\overline N}_a=(1-p_2)p \gamma$. Now every one of these
additional edges has the effect of decreasing the average
distance of the graph by at least $M^{-1}$. Thus, averaging over
all graphs in ${\mathcal G}_{p_1 p_2}(\gamma)$, the mean average
distance must satisfy $\overline d \leq 2- p_2-\gamma p (1-p_2)=
2-p_1 \gamma-p_2 (1-\gamma)$. This, together with the upper
bound, implies that, in mean, $\overline d = 2-p_1 \gamma-p_2
(1-\gamma)$ for large $N$. The fact that this value is a lower
bound for general graphs ensures that, for large values of $N$,
almost every graph has an average distance equal to the mean.

Notice that for graphs in ${\mathcal G}_{p_1}(\gamma)$, in the
infinite $N$ limit, the average distance for $p_1<1$ is only a
function of $\gamma$: ${\overline d}=2-\gamma$. For $p_1=1$ and
$p_2=0$, the graph is circulant, and it is not difficult to see
that the average length is $(1+\gamma)/2 \gamma$ ~\cite{watts}. In
figure~\ref{fig.2} it can be seen that for finite but large
values of $N$, the average distance falls very rapidly with the
number of shortcuts.

\subsection{Clustering coefficient} \label{sec:clust}
The average distance can be thought of as a {\it global} property
of a graph: it gives an idea of how far apart is any vertex from
any other. The {\it clustering coefficient} provides a different
kind of information: it is a measure of how locally connected is
the graph. It is obtained by calculating, for every vertex of the
graph, the number of links joining points of its neighborhood
divided by the total number of possible links in the
neighborhood, and taking its average over all vertices. This
gives the probability that two neighbors of a vertex are
connected to each other. It can be shown~\cite{NSW} that this is
equivalent to calculating the total number of triangles in the
graph, divided by the total number of paths of length 2
(hereafter called $2$-paths).

\begin{figure}
  \centerline{\epsfig{file=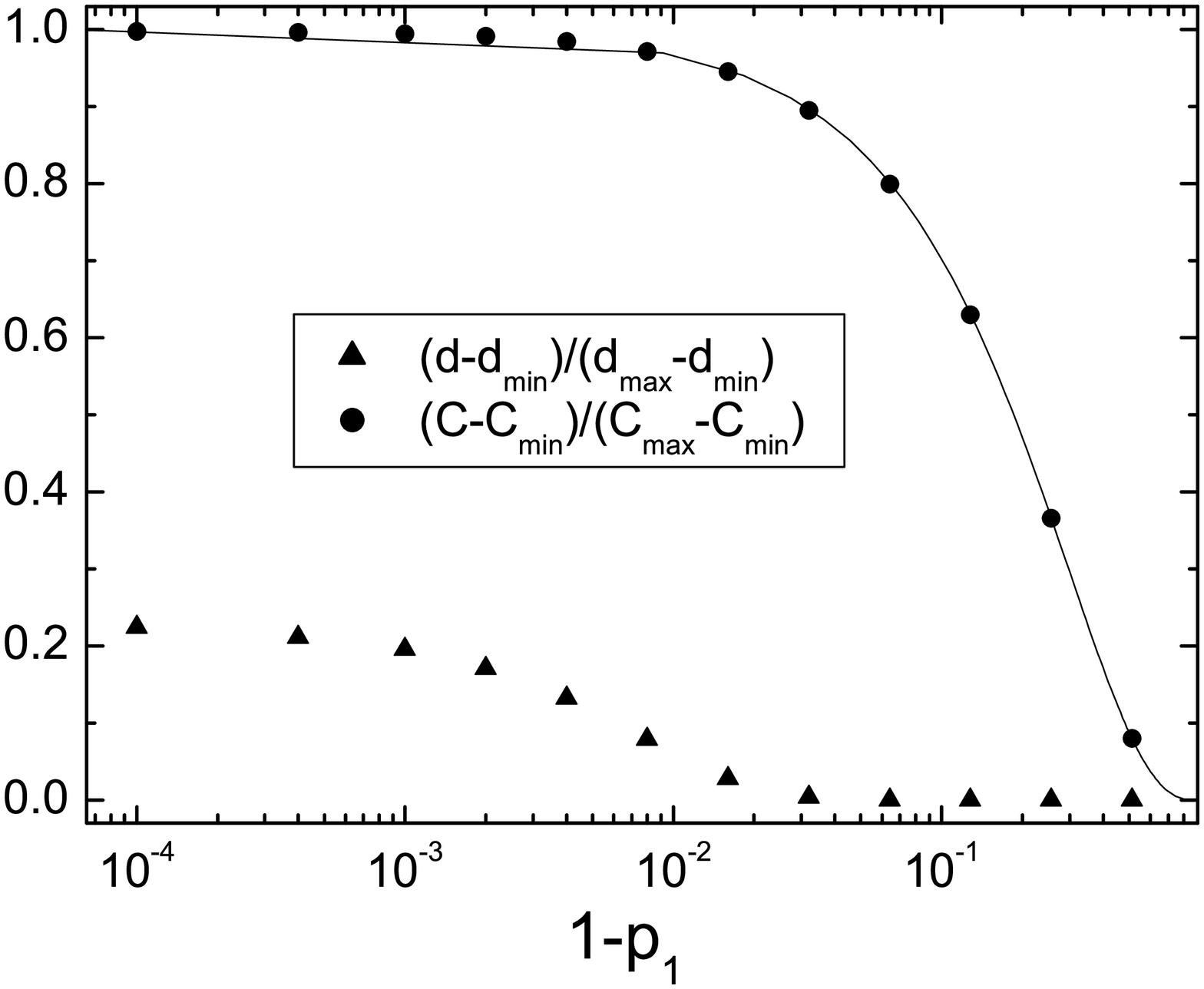,height=7cm}}
  \caption{Average random distance and clustering coefficient for graphs $G_{p_1}(\gamma)$, scaled by
   their maximum and minimum values.
  The symbols are averages
  taken on $30$ matrices with $N=6001$ and $\gamma=0.1$. The line is the theoretical prediction for the
  average distance. Error bars are smaller than the symbols.}
\label{fig.2}
\end{figure}

Instead of calculating this number for every graph and then
average over the ensemble, we calculate an 'annealed' version of
it~\cite{BW}: we take the average of the number of triangles and
divide it by the average number of 2-paths. For large values of
$N$ the agreement with the numerical values turns out to be very
good (see Fig.~\ref{fig.2}).

To calculate the number of 2-paths, one has to find the number of
2-paths for every matrix, weight it with the probability of that
matrix, and then sum over all possible graphs. An equivalent way
of doing this is to sum over all possible paths, with a weight
given by $p_1^i p_2^{2-i}$, where $i$ is the number of long links
in the path.

In the limit of infinite $N$ all vertices are equivalent, and it
is enough to consider all the paths beginning from a fixed
vertex, and then multiply by $N$. To count these paths, one needs
to know the size of the intersection of the $k$-neighborhoods of
two points. If $\gamma<2^{-d}$, this intersection is a connected
region and the calculation of its size is very simple. We analyze
only this case, because one is usually interested in low
dimensions. For bigger values of $\gamma$ the calculations are
straightforward but much more cumbersome.

Assuming the fixed vertex located at the origin, and its other
end at a position $\vec{x} \in Z^d$, the size of the intersection
of the two $k$-neighborhoods is

\begin{equation}
I({\vec x})=\left\{ \begin{array}{ll}
            \prod_{i=1}^{d} I_i({\vec x}) & {\vec x} \in \Gamma_{k}(0) \\
             & \\
            \prod_{i=1}^{d} I_i({\vec x}) -2 & {\vec x} \in \Gamma_{2k}(0)-\Gamma_{k}(0) \\
            & \\
            0 & {\vec x} \in \Gamma_{2k}(0)
            \end{array}
            \right.
\label{eq.int}
\end{equation}

\noindent where $\Gamma_{k}(0)$ is the set of nodes that form the
$k$-neighbourhood of the origin, and

\begin{equation}
I_i({\vec x})=2k_i-|x_i|+1
\end{equation}

Using this, we obtain that the total number of 2-paths is, to first order in $N$,

\begin{eqnarray}
N_p&=& \sum_{{\vec x} \in G} \Pi({\vec x}) [p_1^{2} I({\vec x}) +
2 p_1 p_2 (|\Gamma_k|-I({\vec x}) )+ \nonumber \\&& p_2^2
(N-2|\Gamma_k|+I({\vec x}))] \label{eq.paths}
\end{eqnarray}

\noindent with $\Pi({\vec x})=1$. The first term in the square
bracket counts the number of paths with two short edges, the
second counts the paths with one short and one long edge and the
third counts those paths with two long edges. To calculate the
number of triangles one proceeds in exactly the same way, but it
must now be assumed that the endpoints of the paths are connected
by an edge. Thus, the number of triangles is given by
Eq.~(\ref{eq.paths}) using $\Pi({\vec x})=1-p_1$ for ${\vec x} \in
\Gamma_k(0)$ and $\Pi({\vec x})=1-p_2$ for ${\vec x} \in
V-\Gamma_k(0)$. After some algebra the clustering coefficient
obtained is

\begin{equation}
C=\frac{\gamma_d^2(p_1-p_2)^3((3/4)^d-\gamma_d)+((p_1-p_2)\gamma_d+p_2)^3}{((p_1-p_2)\gamma_d+p_2)^2}
\label{eq.clust}
\end{equation}

\noindent with

\begin{equation}
\gamma_d=\prod_{i=1}^{d} \gamma_i.
\end{equation}

In the case of a graph $G_{p_1}(\gamma)$ the clustering
coefficient simplifies to

\begin{equation}
 C=(p_1-p_2)^3((3/4)^d-\gamma_d)+\gamma_d
 \label{eq.clust2}
\end{equation}

A comparison with data obtained from graphs with 6001 vertices can
be seen in Fig~\ref{fig.2}. Theory and data only differ where
$1-p_1 \simeq (\gamma N)^{-1}$, where our approximations do not
hold.

\section{Spectrum of the graph} \label{sec:spectrum}
For a graph with no loops and with no multiple edges, as those
treated in this article, the adjacency matrix $A$ is a 0-1 matrix
where $A_{ij}=1$ if there is an edge connecting vertices $i$ and
$j$, and $0$ otherwise. Diagonal elements are set to $0$. The
eigenvalues of $A$ provide a lot of information about the
structure of the graph (see~\cite{biggs}). For example, if the
graph is regular and its eigenvalue distribution is symmetric,
then the graph is bipartite. Eigenvalues also provide useful
bounds to quantities as different as the diameter of a
graph~\cite{chung} and the mixing rate~\cite{lovasz} of random
walks. In the following section we summarize some results for the
spectrum of circulant graphs which will be useful in our
determination of the spectrum of DPRGs.

\subsection{Circulant graphs}
The family of circulant graphs is one of the few for which all
the eigenvalues and eigenvectors can be calculated
analytically~\cite{gray}. The adjacency matrix of such a graph is
a {\it circulant} matrix, where row $i$ is simply the first row
shifted $i$ places to the right. In our case, the adjacency
matrices of the graphs with $p_1=1$ and $p_2=0$ are circulant.
They are symmetric matrices with null diagonal elements,
satisfying, for $d=1$ and $j>i$, $A_{ij}=1$ for $j-i \leq k/2$,
and $0$ otherwise.

The eigenvectors of this matrices are $N$ vectors ${\vec
v_j}=(1,\rho_j,\rho_j^2,...,\rho_j^{N-1})$ for $0 \leq j \leq
N-1$, with $\rho_j=\exp{(2 \pi j/N)}$. Interestingly, all
circulant matrices have the same eigenvectors. This is not the
case for the eigenvalues, which satisfy $\lambda^c_j=\sum_{i=1}^N
A_{ji} \rho_j^{i-1}$. In our case, this gives,

\begin{eqnarray}
\lambda^c_j&=&\rho_j^{-(k/2+1)} \sum_{i=1}^N e^{i\frac{\pi k i}{N}}=
\nonumber
\\
&=&2 \cos(\frac{\pi j (k/2+1)}{N}) \frac{\sin(\pi j
k/2N)}{\sin(\pi j/N)} \label{eq.lambj}
\end{eqnarray}

Except for the first one, the Perron-Frobenius eigenvalue
$\lambda^c_0=k$, all the other eigenvalues satisfy
$\lambda^c_j=\lambda_{N-j}$ (we consider only odd values of $N$).
Thus, there are $(N-1)/2$ eigenvalues with multiplicity equal to
2. For $j \ll N$ we have,

\begin{equation}
\frac{\lambda^c_j}{N}=\frac{\sin{(\pi j \gamma)}}{\pi j} -
2\frac{\sin^2(\pi j \gamma/2)}{N}-\frac{\pi j \sin(\pi j
\gamma)}{3 N^2} + O \left(\frac{j^3}{N^4} \right)
\label{eq.lambjapp}
\end{equation}

For $j=O(N)$, it can be seen from Eq.~(\ref{eq.lambj}) that
$\lambda^c_j=O(1)$.

For $d>1$ our substrates are reticles where the range of
connection of each node depends on the spatial direction $i$,
through a variable $k_i$. The adjacency matrices for these graphs
are simply the Kronecker product (or tensor product) of the
corresponding $N^{1/d} \times N^{1/d}$ matrices for each
direction: $A=A(k_1) \bigotimes A(k_2) \bigotimes \dots
\bigotimes A(k_N)$. The resulting matrix is circulant, and has
the nice property that its eigenvalues are

\begin{equation}
\lambda^c_{j_1 j_2 \dots j_d}=\prod_{i=1}^d \lambda^c_{j_i}
\,\,\,\ \mbox{for} \,\,\,\, 1 \leq j_i \leq N \label{eq.lambjd}
\end{equation}

\noindent where the $\lambda^c_{j_i}$s are given by
Eq.~(\ref{eq.lambj}).

\subsection{Dense partially random graphs}
\label{sec:dprg}

The moments $M_j$ of an $N \times N$ matrix $A$ are defined by

\begin{equation}
M_j=N^{-1}Tr(A^j)=N^{-1}\sum_{i=0}^N \lambda_i^j
\label{eq.momj}
\end{equation}

The $j$th power of an adjacency matrix gives information about all
the possible paths of length $j$ in the associated graph. More
specifically, $(A^j)_{ik}$ is the number of j-paths that connect
vertices $i$ and $k$. Thus, the trace of the $j$th power of $A$
is the total number of closed paths, called $cycles$, of length
$j$.

If, for an ensemble of matrices, the calculation of the average
number of cycles can be performed for every length, by using the
averaged version of Eq.~(\ref{eq.momj}) it is possible to obtain
the average distribution of eigenvalues. This is the route taken
by Wigner~\cite{wigner} in his famous derivation of the
semicircle law for random matrices. His work was afterwards
extended~\cite{arnold} to prove that for large matrices, the
distribution of eigenvalues of {\it almost every} matrix of the
ensemble tends to the semicircle. In Ref.~\cite{BG} the same
method was used to calculate the moments of random 0-1 matrices.

In both these works, and in many others, all the cycles are
characterized as sequences of vertices. For each sequence of
length $j$ the statistical weight is the same, as all the edges
in the graph have the same probability of being present.

In our model this is different because short and long links have
different probabilities. Therefore, in our case it is better to
characterize each cycle as a succession of {\it distances}. We
begin by analizing the unidimensional case, and afterwards we
indicate the modifications necessary to extend the results to
higher dimensions.

On a ring, vertices can be consecutively labeled with numbers
from $1$ to $N$ by going round the circle. When standing on a
vertex, a walker can only make a step to the right or to the
left. The distance between vertices $i$ and $j$ is defined as
$d_{ij}=\min(|i-j|,N-|i-j|)$. Intuitively, it is the shortest
number of consecutive vertices, including the end vertex, that a
walker must traverse to go from $i$ to $j$. To define a direction
we say that if a walker, using an existing link, goes from vertex
$i$ to vertex $j$ such that $(i-j)\mod N \leq (N-1)/2$ then it has
made a step to the {\it right} covering a distance $d_{ij}$.
Otherwise, we say that he has traveled to the left, covering a
distance $d_{ij}$. Thus, a $j$-path can be defined by a
succession of $j$ distances. The total distance traveled  is the
sum of the signed distances.

If the path is a closed one, i. e. a cycle, the total signed
distance must be equal to $mN$, because the cycle can contain $m$
complete rounds of the circle. In a cycle traversing $i$ long
edges, called a $ji$-cycle, the number of complete rounds will be
bounded by $m_{max}=\lfloor (j-i) \frac{\gamma}{2}+i(N+1)/(2N)
\rfloor$, where $\lfloor x \rfloor$ gives the largest integer
smaller than or equal to x. $P_{ji}$ is defined as the number of
$ji$-cycles for a fixed position in the cycle of the $i$ long
links. Because the probabilities for each edge are independent,
$P_{ji}$ does not depend on the actual positions of the long
links. As the probability of a $ji$-cycle is $p_1^{j-i} p_2^i$,
the average over ${\mathcal G}_{p_1 p_2}(\gamma)$ of the $j$th
moment is

\begin{equation}
{\overline M}_j=\sum_{i=0}^j \binom{j}{i} p_1^{j-i} p_2^i
{\overline P}_{ji} \label{eq.avMj}
\end{equation}

As explained in the introduction, all the quantities calculated
in this article are averages. Thus, hereafter we drop the
overlines to avoid overloading the notation.

Let us call $d_l$ the distance traveled in the $l$th step. To long
links there correspond signed distances satisfying $k<|d_l| \leq
N/2$. For short links, the corresponding distances satisfy $1
\leq |d_l| \leq k$. Thus, $P_{ji}$ is the number of solutions
$\vec d \in {\textbf Z}^j$ of the problem

\begin{eqnarray}
\sum_{l=1}^j d_l=mN\,\,\,\,\,\,\mbox{for }|m|<m_{max} \nonumber \\
1 \leq |d_l| \leq k/2 \,\,\,\,\,\,\mbox{if }0 \leq l \leq i \nonumber \\
k/2 <|d_l| \leq (N-1)/2\,\,\,\,\,\,\mbox{if }i < l \leq j
\end{eqnarray}

Using the principle of inclusion-exclusion~\cite{vanlint},
$P_{ji}$ can be written as

\begin{equation}
P_{ji}=\sum_{l=0}^i (-1)^{i-l} \binom{i}{l} \tilde{P}_{jl},
\label{eq.Pji}
\end{equation}

\noindent where $\tilde{P}_{jl}$ is the number of solutions to the
simpler problem

\begin{eqnarray}
\sum_{l=1}^j d_l=mN\,\,\,\,\,\,\mbox{for }|m|<m_{max} \nonumber \\
1 \leq |d_l| \leq k/2 \,\,\,\,\,\,\mbox{if }0 \leq l \leq i \nonumber \\
1 <|d_l| \leq (N-1)/2\,\,\,\,\,\,\mbox{if }i < l \leq j.
\label{eq.prob2}
\end{eqnarray}

$\tilde{P}_{ji}$ corresponds to the number of cycles where $j-i$
fixed steps use short links, whereas the other steps can use
either long or short links. It is not difficult to see that
$\tilde {P}_{ji}$ is of order $N^{j-1}$ (as the path is closed,
only $j-1$ steps may be freely chosen). Notice that $d_l$ cannot
be zero because loops are not allowed. But the presence of loops
only adds to $P_{ji}$ terms of order $jN^{j-2}$. As we are only
interested in the dominating term, we include the loop terms,
that allow for simpler calculations. Using this, and rescaling
the distances, the problem of Eq.~(\ref{eq.prob2}) can be
rewritten as the number of solutions of

\begin{eqnarray}
\sum_{l=1}^j d_l=j(l/2+1)+(j-i)(N+1)/2+mN \nonumber \\
\,\,\,\,\,\,\,\,\,\mbox{for }|m|<m_{max} \nonumber \\
1 \leq d_l \leq k+1 \,\,\,\,\,\,\mbox{if }0 \leq l \leq i \nonumber \\
1 \leq d_l \leq N\,\,\,\,\,\,\mbox{if }i < l \leq j.
\end{eqnarray}

Using again the principle of inclusion-exclusion, we get

\begin{eqnarray}
\tilde{P}_{ji}&=&\sum_{m=-m_{max}}^{m_{max}} \sum_{l=0}^j (-1)^l
\sum_{p=0}^{min(l,j-i)} \binom{j-i}{p} \binom{i}{l-p} \cdot
\nonumber \\ && \binom{N \, f_N(p,l,m)+j-1}{j-1}
\Theta(f_N(p,l,m)), \label{eq.tildePji1}
\end{eqnarray}

\noindent where $\Theta(x)$ is the Heaviside (or step)
function($\Theta(x)=1$ for $x>0$, and $\Theta(x)=0$ otherwise) and

\begin{equation}
f_N(p,l,m)=\frac{i}{2}-m+p-l+\gamma(\frac{j-i}{2}-p)-\frac{p+i/2}{N}
\end{equation}

As we are only interested in the dominant terms of $\tilde P_{ij}$, we develop it in powers of $N$ to get

\begin{eqnarray}
\lefteqn{ \tilde{P}_{ji}= N^{j-1} \sum_{m=-m_{max}}^{m_{max}}
\sum_{l=0}^j (-1)^l \sum_{p=0}^{min(l,j-i)} \binom{j-i}{p}
\binom{i}{l-p} \cdot}
\nonumber \\
 && \left(
\frac{f_{\infty}^{j-1}}{(j-1)!}+N^{-1}\frac{f_{\infty}^{j-2}}{(j-2)!}(\frac{j-i}{2}-p)+O(N^{-2})\right)
\Theta(f_{\infty}) \nonumber \\
 &&
\label{eq.tildePji}
\end{eqnarray}

\noindent where

\begin{equation}
f_{\infty}=f_{\infty}(p,l,v)=\frac{i}{2}-m+p-l+\gamma(\frac{j-i}{2}-p).
\end{equation}

For the first order term of Eq.~(\ref{eq.tildePji}), the sums can
be performed (see Appendix) to obtain the simple result

\begin{equation}
\tilde{P}_{ji} \sim N^{j-1} \gamma^{j-i} \,\,\,\,\mbox{ for}
\,\,\, 1 \leq i \leq j \label{eq.tPjisimp}
\end{equation}

For $\tilde{P}_{j0}$ there is not such a simple expression. On the
other hand, $\tilde{P}_{j0}={P}_{j0}$. Thus, $\tilde{P}_{j0}$ is
the number of $j$-cycles of a graph with only short links, that
is, a circulant graph. Thus, it can be written as

\begin{equation}
\tilde{P}_{j0}=P_{j0}=\sum_{l=0}^{N-1} \lambda_l^j
\end{equation}

\noindent where the $\lambda$s are the eigenvalues of a circulant
matrix, given by Eq.~(\ref{eq.lambj}).

Using all this, the number of $ji$-cycles can be calculated,
giving

\begin{equation}
P_{ji}=(-1)^i (P_{j0}-N^{j-1} \gamma^j)+N^{j-1} \gamma^{j-i}
(1-\gamma)^i
\end{equation}

Using this and Eq.~(\ref{eq.avMj}), we obtain, for the first
order of the moments of a dense partially random matrix:

\begin{equation}
M_j=(P_{j0}-\gamma^j N^{j-1}) (p_1-p_2)^j+\gamma_*^j
N^{j-1}\,\,\,\, \mbox{ for }j \geq 3 \label{eq.MjconP}
\end{equation}

\noindent where $\gamma_*=p_1 \gamma + p_2 (1-\gamma)$ is the
average degree of a vertex. The spectrum of eigenvalues that
generates the moments corresponding to Eq.~(\ref{eq.MjconP}),
{\it for all values of $j$} (i. e. not only for $j \geq 3$),
satisfies $\lambda_1/N \rightarrow \gamma_*$ and $\lambda_j/N
\rightarrow (p_1-p_2) \lambda^c_j/N$ for $j \geq 1$. Notice that
this would imply that the whole spectrum is only a rescaling of
the spectrum of the substrate (only the first eigenvalue is
scaling differently). But the problem is that, because of the
symmetry of the matrices, the second moment is, to first order,
$M_2=N \gamma_*$, clearly different to what would be obtained by
setting $j=2$ in Eq.~(\ref{eq.MjconP})
($M_{j=2}=N((\gamma-\gamma^2)(p_1-p_2)^2+\gamma_*))$. This means
that not all the eigenvalues can tend to the values mentioned
above.

On the other hand, the fact that $M_j=O(N^{j-1})$ implies (see
Eq.~(\ref{eq.momj})) that every eigenvalue that satisfies
$|\lambda|=O(N)$ must tend to $(p_1-p_2)\lambda^c$. And from
Eq.~(\ref{eq.lambjapp}) it can be seen that the number of such
eigenvalues diverges with $N$. But the number of eigenvalues that
{\it do not} tend to $\lambda_c$, which satisfy  $|\lambda|=o(N)$,
must also diverge with $N$; otherwise, its influence would not be
felt in $M_2$.

\begin{figure}
  \centerline{\epsfig{file=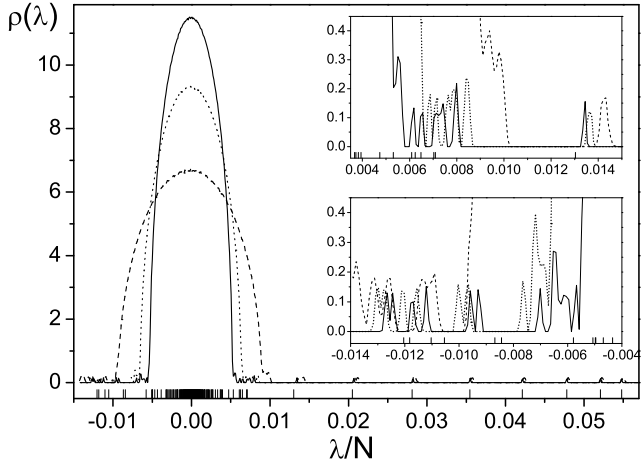,height=7cm}}
  \caption{Average distribution of eigenvalues for $G_{p_1}(\gamma)$ with $\gamma=0.1$ and $p_1=0.6$, for $N=3001$ (dashed line), $N=6001$ (dotted line)
  and $N=9001$ (full line). The averages were taken over $100$, $30$, $10$ matrices respectively.
  The small vertical bars over the horizontal axis show the first order predictions for the discrete part of the spectrum, extended to all values of $\lambda$.}
  \label{fig.3}
\end{figure}

In Fig.~\ref{fig.3} we show the distribution of eigenvalues for
matrices with $N=6001$. It can be seen that the largest
eigenvalues are very close to the values predicted, and the
deviations are larger for smaller values of $\lambda/N$. But for
small values of $\lambda$ the spectrum seems to be continuous and
similar to a semicircle distribution.

If we {\it assume} that the distribution is given by a discrete
part, where to dominant order $\lambda_j \sim \lambda^c_j$ and a
continuous part given by a semicircle distribution (to dominant
order), the second moment can be used to determine the width of
the semicircle. It is known~\cite{wigner} that the semicircle
distribution

\begin{equation}
\rho_s(\lambda)=\left\{ \begin{array}{ll}
\frac{2}{\pi \sigma^2} \sqrt{\sigma^2-\lambda^2} & \mbox{for    }-\sigma \leq \lambda \leq \sigma  \\
\\ &
\\
 0 & \mbox{otherwise    }
\end{array}
\right. \label{eq.semic}
\end{equation}

\noindent generates moments

\begin{equation}
M^s_{2j}=\int d \lambda \, \lambda^{2j} \rho_s(\lambda)=
\frac{2j!}{j!(j+1)!} (\sigma/2)^{2j}
\end{equation}

\noindent (odds moment vanish because of the symmetry.) From
Eq.~(\ref{eq.MjconP}), extended to $j=2$, we know that to obtain
the correct value of the second moment for the whole
distribution, the continuous part should satisfy
$M_2^c=N(\gamma_*-\gamma_*^2-(\gamma-\gamma^2)(p_1-p_2)^2)$. For
this, the width of the semicircle must be

\begin{equation}
\sigma=2
\sqrt{N(\gamma_*-\gamma_*^2-(\gamma-\gamma^2)(p_1-p_2)^2)}
\end{equation}

\begin{figure}
  \centerline{\epsfig{file=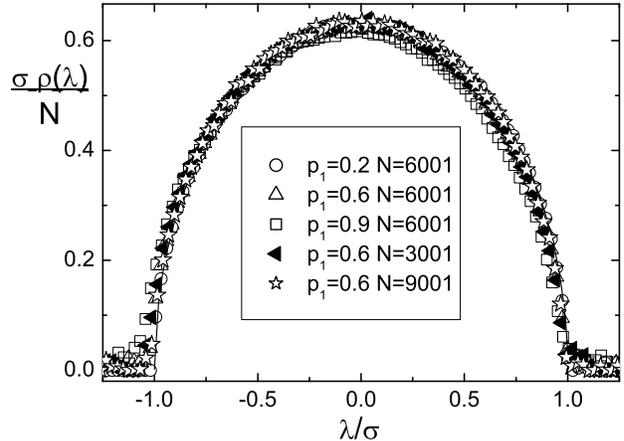,height=7cm}}
  \caption{Scaling of the average of the continuous part of the spectrum for $G_{p_1}(\gamma)$ with $\gamma=0.1$, for different values of $p_1$ and $N$.
  The averages for $N=3001$, $N=6001$ and $N=9001$ have been taken over $100$, $30$ and $10$ matrices, respectively. The full line shows the theoretical prediction (semicircle distribution). }
\label{fig.4}
\end{figure}

To test the correctness of this assumption, we have used it to
scale the average of the continuous part of the spectrum of
$G_{p_1}(\gamma)$ for several values of $N$ and $p_1$. The
results, displayed in Fig.~\ref{fig.4}, show that this scaling
collapses all the curves onto the unit semicircle, as predicted.
Notice that the deviation from the semicircle is only noticeable
for large values of $p_1$, because in this case the discrete part
of the spectrum contains a significant part of the eigenvalues,
thus partially depleting and skewing the semicircle.

To confirm analytically the existence of the semicircle, one
should develop the moments to an order large enough in the
corrections to see the contribution of the semicircle. The problem
is that, whereas the order of magnitude of $M_j^d$ is $N^{j-1}$,
the semicircle generates moments $M_j^s=O(N^{j/2})$. Thus, for
high values of $j$ the contribution of the semicircle lies deeply
buried under the ones of the discrete part of the spectrum.
Nevertheless, we have calculated the moments to the second order
(see next subsection) and confirmed that the semicircle provides
the right value for $M_4$.

The existence of the semicircle imposes a limit on the number of
eigenvalues in the discrete part: it must contain only the
eigenvalues that satisfy $|\lambda|>\sigma$. Eq.~(\ref{eq.lambjapp})
implies that the number of such eigenvalues is proportional to
$N/\sigma=O({\sqrt N})$.

Putting all this together the complete eigenvalue distribution of
DPRGs is, to first order in $N$,

\begin{equation}
\rho(\lambda)=\left\{ \begin{array}{ll}
            \frac{2(N-N_{\sigma})}{\pi \sigma^2N} \sqrt{\sigma^2-\lambda^2} & -\sigma \leq \lambda \leq
            \sigma \\
             & \\
            \frac{2}{N} \sum_{j=0}^\infty \delta(\lambda-N(p_1-p_2) \frac{\sin ( \pi j \gamma) }{ \pi j}) &
            \mbox{otherwise}
            \end{array}
            \right.
\label{eq.distDPRG}
\end{equation}

\noindent where $N_{\sigma}$ is the number of values of $j$ such
that $N(p_1-p_2) \frac{\sin ( \pi j \gamma)) }{ \pi j}>\sigma$,
i. e. the number of eigenvalues contained in the discrete part.
It must be remarked that even though these eigenvalues are
degenerated this degeneracy breaks down for finite values of $N$.
Nevertheless in our simulations their separation is so small as
to make them statistically indistinguishable inside each peak,
for the values of $N$ chosen.

\subsubsection{Second order calculation}
\label{sec:2ndorder}

To go beyond the dominant order in the calculation of the
moments, two different contributions must be taken into account.

The first contribution is simply a refinement of the calculation
of $\tilde P_{ji}$ for $i>0$~\footnote{Although the correction
for ${\tilde P}_{j0}$ can in principle be calculated, we do not
care about the specific functional form of it, because the
correction to the eigenvalues that give rise to this term can be
directly obtained from Eq.~(\ref{eq.lambjapp}).}, which can be
split in two terms. One of them is simply what one gets when
using the second order term from the development of $\tilde
P_{ji}$ (see Eq.~(\ref{eq.tildePji})). This gives $\tilde
P_{ji}^{(2)}=i {\tilde P}_{j-1 \, i-1}$. The other part arises
when we subtract from Eq.~(\ref{eq.tildePji1}) the paths with
loops, i. e, the solutions of Eq.~(\ref{eq.prob2}) having $d_l=0$
for some values of $l$ (that were introduced to make calculations
easier). The number of such solutions is $O(N^{j-l-1})$. Thus, for
the second order calculation we need to subtract from
Eq.~(\ref{eq.tildePji1}) the number of solutions with only one
vanishing distance, which is $iP_{j-1 \, i-1}+(j-i)P_{j-1 \, i}$.
Adding these two terms, and using Eqs.~(\ref{eq.Pji})
and~(\ref{eq.avMj}), the first contribution to the correction of
the moments is obtained:

\begin{equation}
\Delta M_j^{I}=\Delta P_{j0}(p_1-p_2)^j-p_2j M_{j-1}
\end{equation}

\noindent where $\Delta P_{j0}$ is the first correction to
$P_{j0}$.

\begin{figure}
  \centerline{\epsfig{file=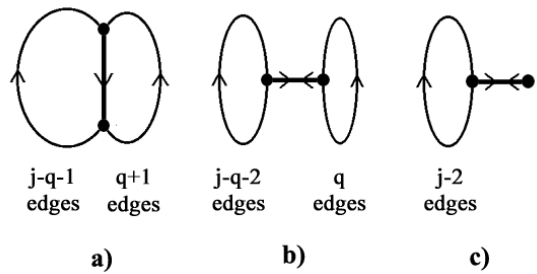, width=7cm}}
  \caption{Schematic representation of the possible cycles with only one repeated edge. The legends indicate the number of edges present in each subcycle. The edge traversed twice is shown as a thick line.}
\label{fig.5}
\end{figure}

The second contribution originates in the fact that the weight
assigned to each cycle in Eq.~(\ref{eq.avMj}), $p_1^{j-i} p_2^i$,
implies that all the edges traversed in each cycle are different.
To correct this, some paths have to be reweighted. But the number
of $j$-cycles with $r$ edges repeated is $O(N^{j-1-r})$ or
smaller, thus we only need to consider cycles with one edge
repeated. But there are only two possible classes of such cycles,
as shown in Fig.~\ref{fig.5}: the first class consists of two
cycles sharing an edge which is traversed in only one direction,
and the second class consists of two cycles joined by an edge
which is traversed in both directions. For the first class
(Fig.~\ref{fig.5}a), the number of cycles is
$O(N^{j-q-2})O(N^{q-1})=O(N^{j-3})$ for all possible values of
$q$. Notice that in this case $q$ must be positive for the edge
to be traversed twice. The number of possible j-cycles for the
second class (Fig.~\ref{fig.5}b) is proportional to $O(N^{j-q-3})
\cdot N \cdot O(N^{q-1})=O(N^{q-3})$ if $q \geq 2$ and
$O(N^{j-3})N=O(N^{j-2})$ if $q=0$. Therefore, the contribution of
dominant order, given by those paths with $q=0$ (shown in fig
Fig.~\ref{fig.5}c), can be written

\begin{eqnarray}
\Delta M_j^{II}&=&  \, \Delta \gamma \, j N \, M_{j-2} \,\,\,
\mbox{for }j \geq 5 , \label{eq.dmjII}
\end{eqnarray}

\noindent where $\Delta
\gamma=\gamma(p_1-p_1^2)+(1-\gamma)(p_2-p_2^2)$. $jN$ is the
number of possibilities for the choice of the edge that will be
repeated, the two terms in $\Delta \gamma$ correspond to the
possibility that the repeated term is a short or a long link, and
$M_{j-2}$ is the number of cycles of length $j-2$.

Notice that to obtain Eq.~(\ref{eq.dmjII}) we have only corrected
the weight of the repeated edge, assuming that in the closed part
of the paths considered (corresponding to the ovals in
Fig.~\ref{fig.5}) all edges are different. For $j>4$ this is
correct, because the number of such subcycles needing reweighting
is of second order respect to the total. But this is not so for
$j=4$. In this case, the oval in Fig.~\ref{fig.5}c corresponds to
a single edge traversed twice. Taking this into account, the right
contribution to the correction of $M_4$ is

\begin{equation}
\Delta M_4^{II} = 4 \, N^2 (\gamma_*^2-\gamma_{**}),
\label{eq.dm4II}
\end{equation}

\noindent where $\gamma_{**}=p_1^2 \gamma+p_2^2 (1-\gamma)$.

Adding both contributions, the first correction to $M_j$ can be
written as:

\begin{eqnarray}
\Delta M_4 &=& (p_1-p_2)^3(\Delta
P_{40}(p_1-p_2)-4p_2(P_{30}-N^2 \gamma^3)) \nonumber \\
 && +N^2(2(\gamma_*^2-\gamma_{**}^2)-4p_2 \gamma_*^3) \,\,\,\,\, \nonumber \\
 \\
\Delta M_j &=& j(p_1-p_2)^{j-2}(P_{j-2\,0}-N^{j-1} \gamma^{j-2})\,
\Delta
\gamma \,\, \nonumber \\
 && -(p_1-p_2)^{j-1} jp_2(P_{j-1\,0}-N^{j-2} \gamma^{j-1}) \nonumber \\
 && +j \, N^{j-2}(-p_2 \gamma_*^{j-1}+\gamma_*^{j-2} \, \Delta \gamma) \nonumber \\ && +(p_1-p_2)^j \Delta
P_{j0}
\,\,\,\,\,\,\,\,\,\,\,\,\,\,\,\,\,\,\,\,\,\,\,\,\,\,\,\,\,\,
\mbox{for }j \geq 5 \label{eq.deltaMj}
\end{eqnarray}

Consider now the distribution of eigenvalues that has been
proposed to generate this moments. We must introduce corrections
to it, to account for the calculated corrections to the generated
moments. For $j \geq 4$ only the corrections to the discrete part
of the distribution are needed. In terms of the fist order term
of the development of the eigenvalues, the correction to the
$j$th moment is

\begin{equation}
\Delta M_j=j \sum_{i=0}^N \lambda_i^{j-1} \Delta \lambda_i .
\end{equation}

Replacing this in the left side of Eq.~(\ref{eq.deltaMj}) and
using that the same holds for the corrections to the moments of
the substrate ($\Delta P_{j0}=\Delta M_j^d$), a comparison of
terms in both sides of the resulting equation gives

\begin{equation}
\Delta \lambda_i=\Delta \lambda_i^0-p_2+\lambda_i^{-1} \Delta
\gamma ,
\end{equation}

But, by construction, these corrections to the discrete
eigenvalues generate the right corrections only to the moments of
order larger than the fourth. For the fourth moment this
discrepancy must be bridged by the corresponding moment generated
by the continuous part of the spectrum. But,

\begin{equation}
\Delta M_{j=4}-\Delta M_{4}= 2(\gamma_* - \gamma_{*2})^2
\end{equation}

\noindent which is exactly the fourth moment generated by the
semicircle distribution given in Eq~(\ref{eq.semic}).

\subsection{Higher dimensional substrates}

The results obtained in the preceding sections can be extended to
DPRGs defined on higher dimensional substrates. These substrates
are defined as hypercubic lattices in $d$ dimensions where each
node is connected to a hypercube of $k_1 k_2...k_d$ other nodes.
As already mentioned, we assume that the hypercube is closed, in
the sense that nodes at the boundaries of the hypercube are
considered nearest neighbors of the nodes at the opposite
boundary.

In a $d$ dimensional DPRG a $ji$-cycle can be represented as a
succession of $j$ distances ${\overrightarrow d}_l \in Z^d$.
Componentwise, this can be regarded as the superposition of $d$
unidimensional subcycles. Notice that the number of shortcuts used
in each subcycle is {\it smaller than or equal to} $i$. To build a
$d$-dimensional $ji$-cycle out of unidimensional paths, only the
following condition must be satisfied. Let us call $\{ i \}$ the
set of steps of the $d$-dimensional $ji$-cycle that traverse
shortcuts, and $\{ i_l \}$ its analog for the subcycle in the
$l$th dimension ($1 \leq l \leq d$). The condition is then that
the {\it union} of all the sets $\{ i_l \}$ be equal to $\{ i \}$:
$\bigcup_{l=1}^d \{ i_l \}=\{ i \}$.

Therefore, counting the number of possible $ji$-cycles is
equivalent to counting the number of unidimensional subcycles
satisfying this condition. This can be written as

\begin{equation}
P_{ji}(d)=\prod_{l=1}^d \sum_{i_l=0}^{i-s_l} \binom{i-s_l}{i_l}
(\prod_{q=1}^{l-1} \binom{i_q}{i_{ql}}) P^l_{j \, i_l+s_{ll}}
\end{equation}

\noindent with $s_l=\sum_{q=1}^{l-1} i_q$ and $s_{ll}
=\sum_{q=1}^{l-1} i_{ql}$. $P^l_{ji}$ is the number of $ji$-paths
in the $l$th dimension. Notice that $i_{lq}$ is the size of the
intersection of sets $\{ i_l \}$ and $\{ i_q \}$. The evaluation
of this expression is cumbersome but straightforward, giving

\begin{equation}
P_{ji}(d)=(-1)^i (P_{j0}(d)-N^{j-1} \gamma_d^j)+N^{j-1}
\gamma_d^{j-i} (1-\gamma_d)^i
\end{equation}

\noindent where $P_{j0}(d)=\prod_{l=1}^d P^l_{j0}$, and
$\gamma_d=\prod_{l=1}^d \gamma (l)$.

Using this and Eq.~(\ref{eq.avMj}), we obtain for the first order
of the moments

\begin{equation}
M_j=(P_{j0}(d)-\gamma_d^j N^{j-1}) (p_1-p_2)^j+(p_1 \gamma_d +
p_2(1-\gamma_d))^j N^{j-1}  \label{eq.MjconP2}
\end{equation}

\noindent for $j \geq 3$. Using the same reasoning of
Section~\ref{sec:dprg} we see that the eigenvalues satisfying
$|\lambda|=O(N)$ must tend to those of the substrate, rescaled by
the disorder: $\lambda_{j_1 j_2 \dots j_d} \sim (p_1-p_2)
\lambda_{j_1 j_2 \dots j_d}^c$, where $\lambda^c_{j_1 j_2 \dots
j_d}$ are given by Eq.~(\ref{eq.lambjd}).

In analogy to the one dimensional case, for the smaller
eigenvalues we can conjecture the presence of a continuous
distribution following a semicircle law, because of the
discrepancy between the real second moment and the one generated
by the discrete distribution. Unfortunately, this conjecture
cannot be tested for all values of $d$ in the same way used for
the unidimensional case in Section~\ref{sec:2ndorder}. The reason
of this is that, as the moments are obtained as sums of products
of $d$ unidimensional moments, their development involves powers
of $N^{1/d}$. But the moments generated by the semicircle are
$O(N^{j/2})$. Thus, for $M_4$ its contribution must be searched in
the $d$th correction to the real fourth moment, whose evaluation,
even though straightforward in principle, gets extremely
cumbersome even for small values of $d$.

Fortunately, for most applications one only needs the largest
eigenvalues in absolute value (see Sec.~\ref{sec:appli} for some
examples), which are given by the discrete part of the
distribution. And if a function of all the eigenvalues is needed,
it can always be rewritten as a series involving the moments.

\subsection{Comparisons}
\label{sec:compa}

The limit of small $\gamma$ should give us an idea of the
approximate form of the spectrum for sparse SWGs. In this limit,
Eqs.~(\ref{eq.p1p2}) and (\ref{eq.p2}) give $p_1 \simeq 1-p$ and
$p_2 \simeq \gamma p$. From Eq.~(\ref{eq.distDPRG}), one can see
that, at least for not too large values of $j$, $\lambda_j \sim
p_1 \lambda^c_0$, for small $\gamma$. Thus, the eigenvalues
accumulate at a distance of $p \gamma N$ from the Frobenius-Perron
eigenvalue $\lambda_0=\gamma N$, with a trail of eigenvalues
reaching to the edges of the semicircle. In the small $\gamma$
limit, the width of the semicircle is $\sigma \simeq 2 \sqrt{2N
\gamma (2p-p^2)}$.

For small values of $p$ the eigenvalues accumulate so close to
the Perron-Frobenius eigenvalues that the gap should only be
visible for very large values of $N$. The continuous part of the
distribution, whose width is proportional to $\sqrt p$, gets very
small and contains few eigenvalues, so its shape becomes very
irregular and skewed to the negative side (to retain the
vanishing of the first moment). This picture is very similar to
what can be seen in Figure 3b of Ref.~\cite{farkas}.

If $p$ is not small, the accumulation point is clearly
separated from the Frobenius-Perron eigenvalue. Besides, $\sigma$
can be close to $\sqrt{N \gamma}$, thus including enough eigenvalues to
take a shape close to the semicircle. This shape should also be
skewed. This picture is very similar to what can be seen in
Figure 3c of Ref.~\cite{farkas}.

It is also similar to what was find in Ref.~\cite{monasson}. In
the graphs considered in that article links are {\it added} to a
sparse substrate (i. e. they are not {\it rewired}). The dense
version of this corresponds to taking $p_1=1$ in our model. Even
though it is the spectrum of the laplacian matrix that is
studied, the results can be translated very easily to the
spectrum of the adjacency matrix, because for large sizes the
graphs can be considered regular, in which case the eigenvalues
of both matrices can be related by the formula
$\lambda^L=k-\lambda^A$. It is found that two peaks appear. The
closest to the Perron-Frobenius eigenvalue is separated from it
by a pseudo-gap (i. e. an interval where there are eigenvalues,
but very few of them). This peak is found to be "in quantitative
agreement with the ring spectrum", and can be related to the
accumulation point mentioned above. The other peak, which is very
irregular for small number of shortcuts, can be related to the
continuous part of the spectrum found in DPRGs.

\section{Some applications}
\label{sec:appli}

It is interesting to notice that the average distance and the
clustering coefficient present qualitatively the same behavior as
that seen in sparse SWGs. We can see that for small values of
$p_1$ the graphs obtained have average distances which are close
to those in random graphs, while retaining a clustering
coefficient close to the values present in circulant graphs.
Naturally the range of values spanned by (the logarithm of) both
quantities is much larger in SWGs.

Having the distribution of eigenvalues, or, equivalently, the
expression for all the moments, allows one to calculate, or at
least to bound many processes that can take place in DPRGs.

For regular graphs (i. e. graphs where all vertices have the same
degree), the spectrum of the adjacency matrix can be very simply
related to the spectrum of the {\it laplacian} matrix, defined by
$L=D-A$ where $A$ is the adjacency matrix and $D$ is the degree
matrix (a diagonal matrix such that $d_{ii}$ is the degree of
vertex $i$) and the {\it normal} matrix $N=D^{-1} A$. The
laplacian has very interesting properties and many applications
in physics, specially because it arises in the discretization of
the Laplacian operator~\cite{mohar}. As SWGs are regular, their
laplacian and adjacency matrix eigenvalues are related by
$\lambda^L=k-\lambda^A$. If the eigenvalues are ordered from
small to large, the first eigenvalue is $\lambda_0^L=0$. The
second eigenvalue is probably the most important as it can be
related to a number of properties of processes taking place in
such graphs.

In the following we show a couple of examples where we apply the
results obtained in the preceding sections.

\subsection{Mixing rate}

A random walk on a graph is defined as a Markov chain where the
probability of jumping from vertex $i$ to vertex $j$ is $1/d_i$
if they are connected, and $0$ otherwise~\cite{lovasz}. Several
properties of random walks can be related to the spectrum of a
graph. For every time $t$ there will be a different probability
$P_t(j)$ of finding the walker on a site $j$. A stationary
distribution is defined by the requirement that $P_{t}=P_{t+1}$
for $t > T$, for some $T$. $\pi(j)=d_j/2M$ is a stationary
distribution, and for regular graphs it is unique. It can be
shown that, regardless of their initial state, all the walks tend
to this distribution, provided the graph is connected and not
bipartite.

When the walk has reached the stationary distribution it has
essentially lost all memory of its initial state (or distribution)
and all the vertices are sampled with probability proportional to
their connectivity, which is useful for several algorithms. But
how fast is the convergence to the stationary distribution? One
of the possible measures of this is the {\it mixing rate},
defined as

\begin{equation}
\mu=\limsup_{t \rightarrow \infty} \max_{i,j} |p_{ij}-d_j/M|^{1/t}
\end{equation}

It can be shown~\cite{lovasz} that $\mu=\lambda_L^N$, the largest
nontrivial eigenvalue of the normal matrix. Using
Eqs.~(\ref{eq.distDPRG}) and (\ref{eq.p1p2}) we obtain,

\begin{eqnarray}
\mu=\lambda_L^N&=&(p_1-p_2) \frac{\sin(\pi \gamma)}{\pi \gamma}
\nonumber \\
&=& \left( \gamma-1+\exp \left(-\frac{\gamma p}{1-\gamma} \right)
\right) \frac{\sin(\pi \gamma)}{\pi \gamma^2}
\end{eqnarray}

\noindent where the last equality is valid for a dense SWG. This
shows that for fixed and small $\gamma$ the mixing rate decreases
almost linearly with the disorder. Notice that the fact that the
average distance jumps to its minimal value at $p=0^+$ does not
influence the mixing rate.

\subsection{Synchronization of coupled oscillators}
One of the most interesting processes that can take place on a
network is the collective dynamics of an array of coupled
oscillators. And perhaps the most striking collective state is
that where all the identical oscillators get {\it synchronized}.
Naturally, synchronization is not always possible, it depends on
the specific properties of the oscillators as well as on the
topology of the network. In Ref.~\cite{PC} a very useful
formalism was introduced to study the conditions for the
existence of a stable synchronized phase, for a wide class of
oscillators and couplings. The equations of motion for the $i$th
oscillator in the network are

\begin{equation}
{\bf x}_i={\bf F}({\bf x}_i)+\sigma \sum_{j=1}^{N} L_{ij} {\bf
H}({\bf x}_j)
\end{equation}

\noindent where ${\bf F}$ governs the dynamics of each individual
oscillator, ${\bf H}$ is an arbitrary output function, $\sigma$
gives the strength of the coupling, and $L$ is the laplacian
matrix of the network. It can be shown~\cite{PC} that for a system
of this form, the condition for the existence of a stable
synchronous state reduces to

\begin{equation}
\lambda_L^L/\lambda_S^L < \beta \label{eq.sync}
\end{equation}

\noindent where $\lambda_L^L$ and $\lambda_S^L$ are, respectively,
the largest and the smallest nontrivial eigenvalues of the
laplacian. $\beta$ is a parameter that depends only on the
oscillators and its coupling, and not on the topology. $\beta \in
[5,100]$ for several chaotic oscillators~\cite{BP}.

\begin{figure}
  \centerline{\epsfig{file=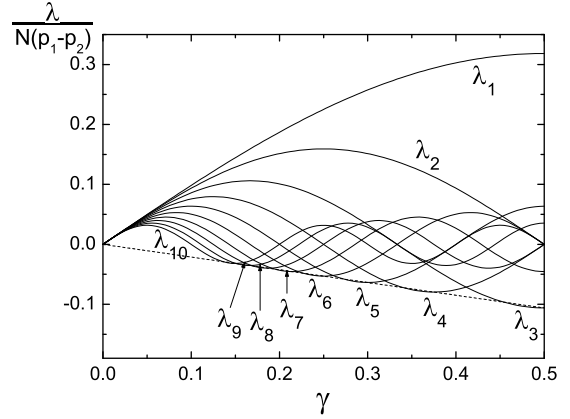,height=6cm}}
  \caption{First ten eigenvalues of the discrete spectrum of the adjacency matrix of a
  DPRG. The dashed line ($\lambda=-2 \gamma N(p_1-p_2)/3 \pi$), that joins the first minima of all the
  eigenvalues, is a reasonably accurate estimation of the smallest
  eigenvalue for every $\gamma$.} \label{fig.6}
\end{figure}

Using Eq.~(\ref{eq.sync}) we can calculate the synchronizability
threshold for dense SWGs. For DPRGs and sufficiently large values
of $N$ the smallest and largest (excluding the Frobenius-Perron)
eigenvalues of the adjacency matrix are located in the discrete
part of the spectrum. The largest nontrivial eigenvalue is always
$\lambda^A_L=N((p_1-p_2)\sin(\pi \gamma)/ \pi$, but the index of
the smallest eigenvalue depends on $\gamma$ (see
Fig.~\ref{fig.6}.) An approximation to it is given by
$\lambda_S^A=-N \gamma (p_1-p_2) 2/3\pi$, which is accurate enough
for our illustrative purposes.

The synchronizability threshold is defined as the smallest value
of $p$ for which the system becomes synchronizable. Using the
already mentioned relation between the adjacency and laplacian
matrices, the synchronization condition for DPRGs becomes

\begin{equation}
\frac{\beta-1}{p_1-p_2} < \frac{2}{3 \pi}+ \frac{\beta
\sin(\gamma \pi)}{\gamma \pi}
\end{equation}

\begin{figure}
  \centerline{\epsfig{file=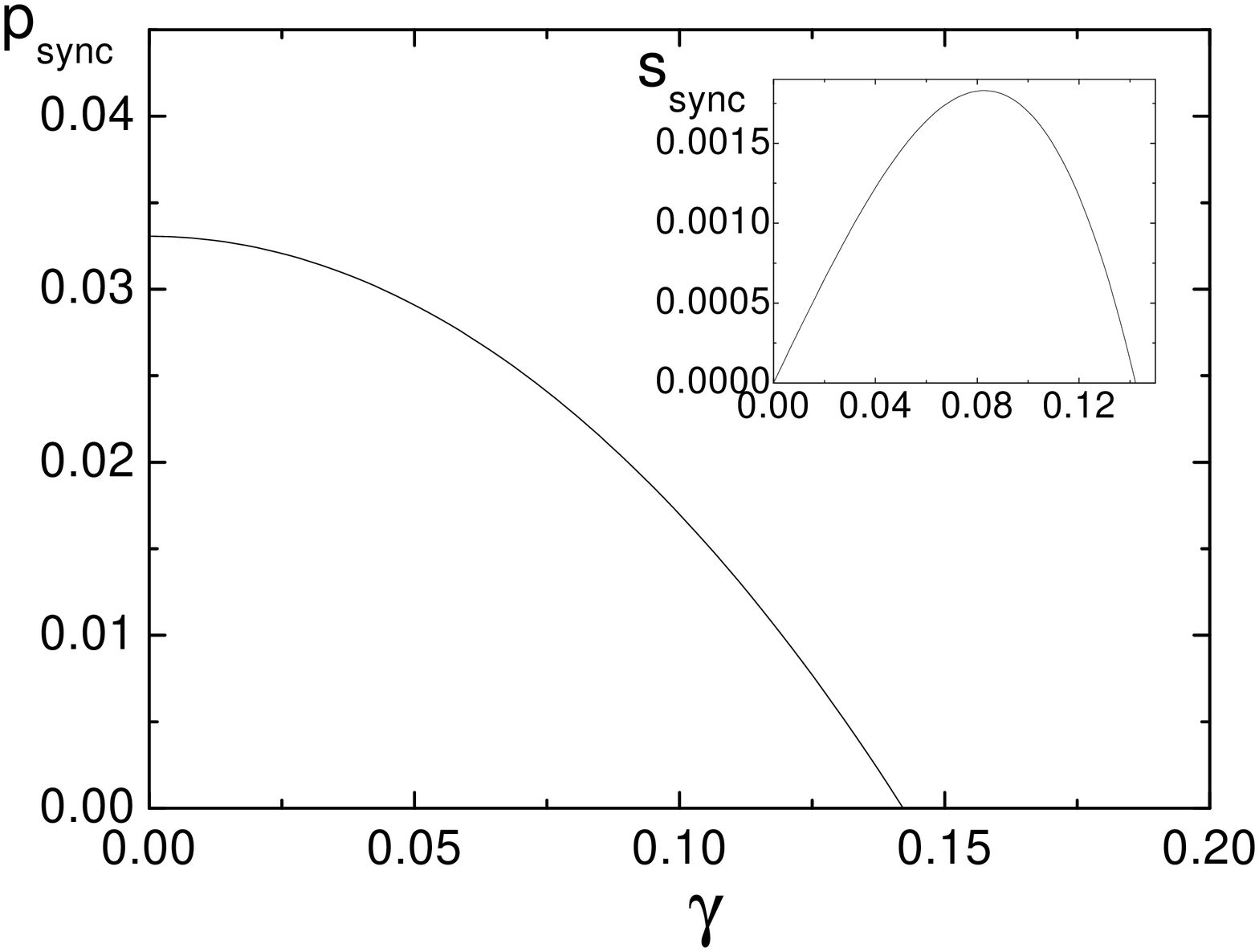,height=6cm}}
  \caption{Synchronization threshold for graphs $G_{p_1}(\gamma)$.
  The inset shows the critical fraction of links that must be
shortcuts. } \label{fig.7}
\end{figure}

Notice that dense random graphs are always synchronizable, as
$\lambda_L/\lambda_S=1$. As was done in Ref.~\cite{BP} we have
calculated the synchronization threshold for a dense SWG for the
case of a system of oscillators with $\beta=37.85$. The results
are displayed in Fig.~\ref{fig.7}.

The region below the curves gives the set of parameters for which
the system is {\it not} synchronizable. It is interesting to
notice that this implies that to obtain a synchronizable system
it is not enough to have a macroscopic number of shortcuts. Also,
the fact that in this region $p$ is positive shows that the onset
of synchronization does not depend on the average distance
because, as we have seen in Section~\ref{sec:avdist}, for $p>0$
the average distance is the smallest possible. This supports the
idea that average distance alone is not a relevant factor for the
onset of synchronization. But in Ref.~\cite{NMLH} it has been
argued that synchronization seems to depend on the combined
effects of small average distance and uniformity of the
connectivity distribution. Yet, our results show that it must at
least depend on some additional factors. For fixed values of
$\gamma$ we have seen that the synchronization threshold is $p_t
> 0$, even though all graphs $G_p (\gamma)$ have the same connectivity ($N
\gamma$ connections per node) and the same average distance for
$p>0$.

\section{Conclusions}
\label{sec:conclu}

In this article we have studied some properties of dense partially
random graphs, that can be obtained by adding edges with probability $p_2$ to a dense ordered substrate, and then deleting the original edges of the substate with probability $1-p_1$. We have found that
they show qualitatively the same behaviour as the corresponding
properties of sparse graphs, in particular small world graphs.
For the average distance we find that, for any macroscopic number
of shortcuts, it falls to its minimum possible value (in the
infinite size limit.) We show that the clustering coefficient
decays slowly, thus allowing for a range of parameters where the
graphs have relatively large clustering and minimal average
distance.

By counting cycles on the graph, we have obtained the distribution
of eigenvalues of the adjacency matrix. We found that it consists
of two parts: a discrete one where the eigenvalues, of order $N$,
are simply rescalings of the corresponding eigenvalues of the
substrate, and a continuous part given by a semicircle
distribution whose width is of order $\sqrt N$, and is
proportional to the disorder. It is interesting to notice that a
similar form of the spectrum has been obtained~\cite{bolla} for
matrices that are the sum of a stochastic matrix and a block
matrix. The block matrix is composed of $k$ blocks of size
proportional to $N$ where all the off diagonal components are
equal. The spectrum obtained consists of a semicircle
distribution of width proportional to $\sqrt N$ and a discrete
part containing the $k$ largest eigenvalues (of order $N$).
Notice that in our case the discrete part contains a diverging
number of eigenvalues.

We have shown how the distribution found can be useful to
understand the distributions that arise in the numerical study of
the spectrum of sparse small world graphs. In the studies
published so far two peaks appear, and a pseudo-gap separates the
bulk of the spectrum from the Frobenius Perron eigenvalue.
Comparing with our results for small values of the connectivity,
one of the peaks can in principle be associated to large
eigenvalues of the substrate, all rescaled by the same value
(dependent on the disorder), and the other to a continuous
distribution of small width that is usually a signature of
disorder.

We have applied our results to the calculation of the mixing rate
of a random walk on the graph, and to the calculation of the
synchronization threshold of a system of coupled oscillators
placed on the nodes of the graph. We have shown that below and
above the threshold there exist graphs with the same average
distance and the same connectivity (i. e. the same number of
connections per vertex). Previously it has been argued that a
combination of these two factors was responsible for the onset of
synchronizability. Our results show that there must at least
exist more quantities involved.

\section*{Acknowledgments}
I acknowledge support from the Centro Latinoamericano de
F\'{i}sica.

\section*{Appendix}
Here we sketch the calculation that leads to
Eq.~(\ref{eq.tPjisimp}). Although we suspect that there must be a
shorter path to Eq.~(\ref{eq.tPjisimp}), we have not been able to
find it.

We begin by splitting the dominant term in
Eq.~(\ref{eq.tildePji}), ${\tilde P}_{ji}^{(1)}$, in two parts:

\begin{eqnarray}
\tilde{P}_{ji}^{(1)}&=& \sum_{m=-m_{max}}^{m_{max}} \sum_{l=0}^j
(-1)^l \sum_{p=0}^{\min(l,j-i)} \binom{j-i}{p}
\binom{i}{l-p} \nonumber \\
 &&\cdot \frac{N^{j-1}}{(j-1)!} g(i/2-m+p-l+ \gamma (\frac{j-i}{2}-p)) \nonumber \\ &=& \frac{N^{j-1}}{(j-1)!} ({\bf A} + {\bf B})
\label{eq.AmasB}
\end{eqnarray}

\noindent where $g(x)=x^{j-1} \theta(x)$ and

\begin{eqnarray}
{\bf A}&=& \sum_{m=-m_{max}}^{m_{max}}
\sum_{l=j-i}^j (-1)^l \sum_{p=0}^{j-i} \binom{j-i}{p}
\binom{i}{l-p} \cdot \nonumber \\
 &&\cdot g(i/2-m+p-l+ \gamma (\frac{j-i}{2}-p)) \\
{\bf B}&=& \sum_{m=-m_{max}}^{m_{max}}
\sum_{l=0}^{j-i-1} (-1)^l \sum_{p=0}^{l} \binom{j-i}{p}
\binom{i}{l-p} \cdot \nonumber \\
 &&\cdot g(i/2-m+p-l+ \gamma (\frac{j-i}{2}-p))
\label{eq.AmasB2}
\end{eqnarray}

Notice that, to avoid overloading the notation, we extend the
definition of the combinatorial numbers,
$\binom{a}{b}~=~\frac{a!}{(a-b)!b!}$ for $b \leq a$, to
$\binom{a}{b}=0$ for $b>a$. By making the replacements $p
\rightarrow j-i-p$ and $l \rightarrow j-i-l$, and rearranging the
sums, we obtain:

\begin{eqnarray}
{\bf A}&=& \sum_{m=-m_{max}}^{m_{max}}
 \sum_{p}^{j-i} \prime \sum_{l=0}^{i-p} (-1)^{l+j-i}
\binom{i}{l+p} (-1)^l \nonumber \\&& \cdot g(i-v-p-l+ \gamma
p)={\bf A'}-{\bf B'}  \label{eq.AmasBp}
\end{eqnarray}

\noindent where $\sum_i^a \prime=\sum_{i=0}^{a}(-1)^i
\binom{a}{i}$ is an operator and

\begin{eqnarray}
{\bf A'}&=& (-1)^j  \sum_p^{j-i}
\prime \sum_l^i \prime \sum_{m=0}^{2m_{max}} g(v+l+\gamma p - \alpha) \\
{\bf B'}&=& (-1)^{j-i} \sum_{m=0}^{2m_{max}} \sum_{p=1}^{j-i}
(-1)^p \binom{j-i}{p} \sum_{l=0}^{p-1} (-1)^l  \binom{i}{l} \cdot \nonumber \\
 &&\cdot g(v+l+\gamma p - \alpha)
\label{eq.ApBp}
\end{eqnarray}

By manipulating the indices and reordering the sums, it is not
difficult to show that ${\bf B}={\bf B'}$. Using the definition of
$\alpha$, we can show that $\lfloor l+\gamma p - \alpha \rfloor <2
m_{max}$. Thus

\begin{eqnarray}
{\bf A'}&=& (-1)^j  \sum_p^{j-i}
\prime \sum_l^i \prime \sum_{m=-l}^{\lfloor \gamma p - \alpha \rfloor} (-m+\gamma p - \alpha)^{j-1} \nonumber \\
&=&   \sum_p^{j-i}
\prime \sum_{l=1}^i (-1)^{l-i} \binom{i}{l} \sum_{m=-l}^1 (-m+\gamma p - \alpha)^{j-1}+ \nonumber\\
&+& \sum_{p=\lceil \alpha / \gamma \rceil}^{j-i}(-1)^{p-j}
\binom{j-i}{p} \sum_{l}^i \prime \nonumber \\
& &\cdot \sum_{m=0}^{\lfloor \gamma p -
\alpha \rfloor} (-m+\gamma p - \alpha)^{j-1} \nonumber \\
\label{eq.Ap2}
\end{eqnarray}

It is known~\cite{vanlint} that

\begin{equation}
\sum_k^a \prime k^n=\left\{ \begin{array}{ll}
            (-1)^a a! S(n,a) &\mbox{for   }0<a \leq n \\
             & \\
            0 &\mbox{for   }a > n
            \end{array}
            \right.
\label{eq.vanlint}
\end{equation}

\noindent where $S(n,a)$ are the Stirling numbers of the second
kind~\cite{vanlint}. Using Eq.~(\ref{eq.vanlint}) one sees that
the term in the last line of Eq.~(\ref{eq.Ap2}) vanishes (but
notice that this can only happen for $i>0$). Thus, expanding in
powers of $m$, we get

\begin{eqnarray}
{\bf A'}&=&  \sum_p^{j-i} \prime \sum_{l=1}^i (-1)^{l-j} \binom{i}{l} \sum_{k=0}^{j-1}
(\gamma p - \alpha)^{j-1-k} \sum_{m=1}^l m^k \nonumber\\
\label{eq.Ap3}
\end{eqnarray}

Bernoulli's expression for a sum of powers~\cite{ireland} is
$\sum_{m=1}^l m^k=\sum_{v=1}^{k+1} b_{km} l^v$ where
$b_{km}=\frac{(-1)^{k-m+1}}{k+1} \binom{p+1}{k} B_{k-m+1}$ and
$B_i$ are the Bernoulli numbers. Using this, and summing over
$l$, we get

\begin{eqnarray}
{\bf A'}&=&  \sum_{k=i-1}^{j-1} \binom{j-1}{k} \sum_{v=i}^{k+1}
b_{vk} (-1)^{v-j} i! S(v,i) \nonumber \\ && \cdot\sum_p^{j-i}
\prime (\gamma p - \alpha)^{j-1-k} \label{eq.Ap4}
\end{eqnarray}

But Eq.~(\ref{eq.vanlint}) shows that in Eq.~(\ref{eq.Ap4}) the
sum over $p$ vanishes for $k>i-1$. Thus, only one term survives,
the one with $k=i-1$. Using that $B_0=1$ and $S(i,i)=i!$ we finally
obtain

\begin{equation}
\tilde{P}_{ji} \sim \frac{N^{j-1}}{(j-1)!} {\bf A'}=N^{j-1}
\gamma^{j-i}
\end{equation}

\end{document}